\documentclass[a4paper,unpublished,onecolumn]{quantumarticle}
\pdfoutput=1

\usepackage[utf8]{inputenc}
\usepackage[english]{babel}
\usepackage[T1]{fontenc}

\usepackage{amsmath}
\usepackage{amssymb}
\usepackage{graphicx}
\usepackage{lipsum}
\usepackage{appendix}
\usepackage{float}
\usepackage{bm}
\usepackage{dsfont}

\usepackage[italicdiff]{physics}
\usepackage[colorlinks=true,allcolors=blue]{hyperref}
\usepackage[caption=false]{subfig}
\usepackage{enumitem}

\allowdisplaybreaks
\newcommand{\unitop}{\hat{\mathds{1}}}

\begin{document}

\title{An Accurate Pentadiagonal Matrix Solution for the Time-Dependent Schr\"{o}dinger Equation}

\author{Ankit Kumar}
\orcid{0000-0003-3639-6468}
\email{kumar.ankit.vyas@gmail.com}

\affil{Department of Physics, Indian Institute of Technology Roorkee, Roorkee 247667, India}

\begin{abstract}
One of the unitary forms of the quantum mechanical time evolution operator is given by Cayley's approximation. A numerical implementation of the same involves the replacement of second derivatives in Hamiltonian with the three-point formula, which leads to a tridiagonal system of linear equations. In this work, we invoke the highly accurate five-point stencil to discretize the wave function onto an Implicit-Explicit pentadiagonal Crank-Nicolson scheme. It is demonstrated that the resultant solutions are significantly more accurate than the standard ones. We also discuss the resolution of bipartite wavepacket dynamics and derive conditions under which a product state from the laboratory perspective remains a product state from the center-of-mass point of view. This has profound applications for decoupling complicated bipartite dynamics into two independent single-particle problems.
\end{abstract}

\maketitle

\setcounter{secnumdepth}{2}
\setcounter{tocdepth}{2}
\tableofcontents

\section{Time evolution in quantum mechanics}

The quantum mechanical state of a particle is described in position representation by a wave function $\psi$. In the non-relativistic limits, this wave function evolves with time in accordance with the time-dependent Schr\"{o}dinger equation (TDSE):
\begin{equation}
 i\hbar\pdv{t}\psi(\bm{r},t) = \qty( -\frac{\hbar^2}{2m}\nabla^2 + V(\hat{\bm{r}},t) ) \psi(\bm{r},t),
 \end{equation}
where $m$ is the mass of the particle, and $V(\hat{\bm{r}},t)$ is the potential. In many physical problems the potential is static, i.e., $V(\hat{\bm{r}},t)=V(\hat{\bm{r}})$, and the resolution of TDSE is equivalent to the implementation of the time-evolution operator $\hat U$:
\begin{equation}
\psi(\bm{r},t+\Delta t) = \hat U(\Delta t) \ \psi(\bm{r},t) = \exp(-i\frac{\Delta t}{\hbar}\hat H) \psi(\bm{r},t),
\end{equation}
where $\hat H = -(\hbar^2/2m)\nabla^2 + V(\hat{\bm{r}})$ is the Hamiltonian. Note that $\hat U$ is unitary, which ensures the norm (total probability) is preserved at all times:
\begin{equation}
\braket{\psi}_{\Delta t} = \ev{\hat U^\dagger \hat U}{\psi}_{0} = \braket{\psi}_{0}.
\end{equation}
In this work we develop an efficient numerical scheme for a precise resolution of single-particle TDSE, and discuss a strategy to utilize the same set of tools to handle the bipartite wave packets in the COM frame of reference.

The complexities in calculating the time dependence of $\psi$ depend on the functional form of the interaction.
Even for simple Gaussians as initial states, closed analytical forms are calculable only in trivial situations, e.g., in the free space~\cite{GaussEvolFreeSpace_SMBlinder}, and the harmonic oscillator potential~\cite{GaussEvolHarmOsc_Tsuru}. Simple harmonic oscillators are regarded as the most precious tools of a theoretical physicist, but none of the fundamental forces in nature behaves so.
This demands an efficient generic numerical scheme to precisely solve the quantum evolution for arbitrary potentials which may be encountered in realistic laboratory conditions.
Along this line, the first step would be to approximate $\hat U$ up to the first order in a series expansion:
\begin{equation}
\hat U (\Delta t) =  \exp(-i\frac{\Delta t}{\hbar}\hat H)  = \sum_{n=0}^{\infty} (-i)^n \frac{\Delta t^n}{\hbar^n} \hat H^n \approx \unitop - i\frac{\Delta t}{\hbar}\hat H.
\end{equation}
However, such truncation leads to a loss of unitarity, which in turn leads to a change in total probability over time:
\begin{eqnarray}
\braket{\psi}_{\Delta t} &=&  \ev{\hat U^\dagger \hat U}{\psi}_0   	\nonumber	\\
&=&  \ev{\qty( \unitop + i\frac{\Delta t}{\hbar}\hat H ) \qty( \unitop - i\frac{\Delta t}{\hbar}\hat H )}{\psi}_0	\nonumber \\
&=&  \ev{\qty( \unitop - i\frac{\Delta t}{\hbar}\hat H + i\frac{\Delta t}{\hbar}\hat H + \frac{\Delta t^2}{\hbar^2} \hat H^2 )}{\psi}_0	\nonumber \\
&=& \braket{\psi}_{0} + \frac{\Delta t^2}{\hbar^2} \ev{\hat H^2}{\psi}_0 > \braket{\psi}_{0} .
\end{eqnarray}
While this is acceptable for short times (the norm is preserved up to the linear order in $\Delta t$), the errors accumulate on (realistic) longer time scales, quickly leading to divergence. Moreover, such approximations do not respect the bidirectional stability in time.
One may be tempted to include higher-order terms in the series expansion, but this would require an impractical numerical evaluation of various higher-order derivatives of the wave function. We must therefore look for alternative ways to integrate the TDSE precisely.

Various techniques have been established that are stable and mitigate errors within their respective capacities~\cite{FEIT1982412,Park1986,BANDRAUK1991428,Muller1999,Nurhuda1999,Watanabe2000}. In this work, we chose to utilize Cayley's form of evolution operator as it circumvents all of our problems with an \emph{unconditional stability} over long time scales~\cite{book_FiniteDiff_JWThomas}. 
We approximate the second-order derivatives with the highly accurate five-point stencil to discretise the problem onto a pentadiagonal Crank-Nicolson scheme. The resultant solutions are much more accurate compared to the standard tridiagonal ones. This will be useful in situations where the potential is very weak, e.g., the gravitational field between two nearby quantum masses.

We thereafter focus on the resolution of the bipartite quantum dynamics, assuming that both the particles are initially prepared in Gaussian wave packets. The usual coordinate transformations to the center of mass (COM) frame of reference are discussed. At least for central interactions, the Hamiltonian decouples into the COM and the relative degrees of freedom, and the product form of a quantum state in this division is maintained at all times. However, a complete decoupling of the dynamics requires the initial quantum state to be separable in the COM frame of reference.
For a two-mode Gaussian state this happens only when the two particles are cooled in the ground state of identical harmonic traps. Note that, unlike regular problems where the COM is described by a plane wave, here it is described by a localised wave packet undergoing proper quantum mechanical time evolution.
The reduced mass wave packet evolves in the interaction sourced from the COM and based on the functional form of the potential. The time evolution can be dealt either analytically or numerically.

\section{Cayley's form of evolution operator}

Cayley's form is a fractional approximation of the quantum mechanical evolution operator. The underlying idea is to evolve $\psi(\bm{r},t)$ by half of the time step forward in time, and $\psi(\bm{r},t+\Delta t)$ by half of the time step backward in time, such that they agree at time $t+\Delta t/2$~\cite{KOSLOFF198335,Ankit_2022_TDSE,Ankit_2022_Gravity,Ankit_2021_Quantum,CoP_PPuschnig}:
\begin{eqnarray}
\ket{\psi}_{t} \xrightarrow[]{\Delta t/2} &\bullet& \xleftarrow[]{\Delta t/2} \ket{\psi}_{t+\Delta t}
\nonumber	\\
\implies \hat U\qty( + \frac{\Delta t}{2} ) \ \psi(\bm{r},t) &=&  \hat U\qty( - \frac{\Delta t}{2} ) \ \psi(\bm{r},t+\Delta t)
\nonumber \\
\implies  \exp(-i\frac{\hat H\Delta t}{2\hbar}) \psi(\bm{r},t) &=&  \exp(+i\frac{\hat H\Delta t}{2\hbar}) \psi(\bm{r},t+\Delta t).
\end{eqnarray}
With a first-order approximation on both sides,
\begin{equation}
\qty(  \unitop- i\frac{ \hat H\Delta t}{2\hbar}  ) \psi(\bm{r},t)
\approx
 \qty(  \unitop+ i\frac{ \hat H\Delta t}{2\hbar}  ) \psi(\bm{r},t+\Delta t) ,
\label{eq:ImpExpExpression}
\end{equation}
we arrive at 
\begin{equation}
 \psi(\bm{r},t+\Delta t) = \qty(  \unitop+ i\frac{ \hat H\Delta t}{2\hbar}  )^{-1} \qty(  \unitop- i\frac{ \hat H\Delta t}{2\hbar}  ) \psi(\bm{r},t).
\end{equation}
Hence, Cayley's form of evolution operator is given by
\begin{equation}
\hat U(\Delta t) = \left( \unitop+ i\frac{ \hat H\Delta t}{2\hbar} \right)^{-1} \left( \unitop- i\frac{ \hat H\Delta t}{2\hbar} \right).
\label{eq:cayley_operator}
\end{equation}
A replacement of the second-order derivatives in Hamiltonian with finite difference formulas tells us that the wave function at different times is related by a Crank-Nicolson (CN) scheme, which is unconditionally stable for TDSE-like problems~\cite{book_FiniteDiff_JWThomas}.

The total probability is preserved over time as the resultant evolution operator in Eq.~\eqref{eq:cayley_operator} is unitary, as shown below. 
The hermitian conjugate of $\hat U$ is
\begin{eqnarray}
	\hat U^\dagger &=&  \qty[ \left( \unitop+ i\frac{ \hat H\Delta t}{2\hbar} \right)^{-1} \left( \unitop- i\frac{ \hat H\Delta t}{2\hbar} \right) ]^\dagger 	\nonumber \\
	&=& \left( \unitop- i\frac{ \hat H\Delta t}{2\hbar} \right)^\dagger \left( \unitop+ i\frac{ \hat H\Delta t}{2\hbar} \right)^{-1,\dagger},	\hspace{1cm}	: \qty{ \qty(\hat A \hat B)^\dagger = \hat{B}^\dagger \hat{A}^\dagger }.
\end{eqnarray}
Note that for any operator $\hat A$ we have
\begin{eqnarray}
&&	\hat{A} \hat{A}^{-1} = \unitop
\nonumber	\\
\implies 	&&	\qty( \hat{A} \hat{A}^{-1} )^\dagger = \unitop^\dagger
\nonumber	\\
\implies 	&&	\qty(\hat{A}^{-1})^\dagger \hat{A} ^\dagger = \unitop,	\hspace{1cm}
 :\qty{ \qty(\hat A \hat B)^\dagger = \hat{B}^\dagger \hat{A}^\dagger },
\nonumber	\\
\implies 	&&	 \qty(\hat{A}^{-1})^\dagger \hat{A} ^\dagger \qty(\hat{A} ^\dagger)^{-1} = \unitop  \qty(\hat{A} ^\dagger)^{-1} 
\nonumber	\\
\implies 	&&	 \qty(\hat{A} ^{-1})^\dagger \unitop = \unitop  \qty(\hat{A} ^\dagger)^{-1}  	
\nonumber	\\
\implies 	&&	 \qty(\hat{A} ^{-1})^\dagger = \qty(\hat{A} ^\dagger)^{-1}.  	
\end{eqnarray}

Accordingly, the hermitian conjugate of $\hat U$ can be re-written as\
\begin{eqnarray}
	&=& \left( \unitop - i\frac{ \hat H\Delta t}{2\hbar} \right)^\dagger \left( \unitop + i\frac{ \hat H\Delta t}{2\hbar} \right)^{\dagger,-1}		\nonumber \\
	&=& \left( \unitop + i\frac{ \hat H\Delta t}{2\hbar} \right) \left( \unitop - i\frac{ \hat H\Delta t}{2\hbar} \right)^{-1},
\end{eqnarray}
which implies
\begin{equation}
\hat U \hat U^\dagger
	= \left( \unitop+ i\frac{ \hat H\Delta t}{2\hbar} \right)^{-1} \left( \unitop- i\frac{ \hat H\Delta t}{2\hbar} \right) \left( \unitop + i\frac{ \hat H\Delta t}{2\hbar} \right) \left( \unitop - i\frac{ \hat H\Delta t}{2\hbar} \right)^{-1}.
\end{equation}
The two terms in the middle commute, and hence
\begin{equation}
	\hat U \hat U^\dagger
\equiv \left( \unitop+ i\frac{ \hat H\Delta t}{2\hbar} \right)^{-1} \left( \unitop + i\frac{ \hat H\Delta t}{2\hbar} \right)  \left( \unitop - i\frac{ \hat H\Delta t}{2\hbar} \right) \left( \unitop - i\frac{ \hat H\Delta t}{2\hbar} \right)^{-1} = \unitop.
\end{equation}

\subsection{The tridiagonal discretisation}

The standard practice for calculating numerical derivatives is to implement various finite-difference approximations. In this work, we only deal with one-dimensional problems, i.e., when
\begin{equation}
\hat H = - \frac{\hbar^2}{2m} \pdv[2]{x} + V(\hat x) .
\end{equation}
The easiest is to replace the second derivative in Hamiltonian with the three-point central-difference formula,
\begin{equation}
f''(x) \approx \frac{f(x+\Delta x) - 2f(x) + f(x-\Delta x)}{\Delta x^2} + \mathcal{O}(\Delta x^2),
\end{equation}
which implies that Eq.~\eqref{eq:ImpExpExpression} is transformed to
\begin{equation}
\begin{split}
\psi_j^{n+1} + \frac{i \Delta t}{2\hbar} \left[ -\frac{\hbar^2}{2m} \left( \frac{\psi_{j+1}^{n+1} - 2\psi_{j}^{n+1} + \psi_{j-1}^{n+1}}{\Delta x^2} \right) + V_j  \psi_j^{n+1} \right] \\
= \psi_j^{n} - \frac{i \Delta t}{2\hbar} \left[ -\frac{\hbar^2}{2m} \left( \frac{\psi_{j+1}^{n} - 2\psi_{j}^{n} + \psi_{j-1}^{n}}{\Delta x^2} \right) + V_j  \psi_j^{n} \right],
\end{split}
\label{TDSE_disc_tridiag}
\end{equation}
where $f_j^n \equiv f(x_j,t_n)$, $\Delta x=x_{j+1}-x_j$ is the grid size, and $\Delta t=t_{n+1}-t_n$ is the time step. To simplify the notation, let us call:
\begin{eqnarray}
\zeta_j^n &=& \psi_j^{n} - \frac{i \Delta t}{2\hbar} \left[ -\frac{\hbar^2}{2m} \left( \frac{\psi_{j+1}^{n} - 2\psi_{j}^{n} + \psi_{j-1}^{n}}{\Delta x^2} \right) + V_j  \psi_j^{n} \right],
\nonumber	\\
a_j &=& 1 + \frac{i \Delta t}{2\hbar} \left( \frac{\hbar^2}{m\Delta x^2} + V_j \right),
\hspace{1cm}
b  =  - \frac{i\hbar\Delta t}{4m\Delta x^2}.
\end{eqnarray}
Eq.~(\ref{TDSE_disc_tridiag}) can now be re-written as
\begin{equation}
\begin{pmatrix}
a_1 & b \\
\ddots & \ddots & \ddots \\
& b & a_{j-1} & b \\
& & b & a_j & b \\
& & & b & a_{j+1} & b \\
& & & & \ddots & \ddots & \ddots \\
& & & & & b & a_{J-1}
\end{pmatrix} \\
\cdot \begin{pmatrix}
\psi_1^{n+1} \\ \vdots \\ \psi_{j-1}^{n+1} \\ \psi_{j}^{n+1} \\ \psi_{j+1}^{n+1} \\ \vdots\\ \psi_{J-1}^{n+1} \end{pmatrix} = \begin{pmatrix}
\zeta_1^{n} \\ \vdots \\ \zeta_{j-1}^{n} \\ \zeta_{j}^{n} \\ \zeta_{j+1}^{n} \\ \vdots \\ \zeta_{J-1}^{n}
\end{pmatrix},
\label{TLSE_matrix}
\end{equation}
where $J$ is the dimension of the position grid. The matrix on the left is composed entirely of constants and the old wave function is stored in the column vector on the right, $\zeta$. 
The standard practice to solve such a system of linear equations is through the Thomas algorithm (which is nothing but Gaussian elimination in a tridiagonal case).

\subsection{The pentadiagonal discretisation}
\label{sec:pentadiagonaldiscretisation}

The three-point formula for the second derivative is accurate up to an error $\mathcal{O}(\Delta x^2)$. The corresponding tridiagonal discretisation works very well in most situations, except for astonishingly weak potentials, e.g., the gravitational field between two quantum particles. In such cases, it falls short due to the accumulation of errors over time. A numerical scheme is as good as the underlying finite-difference approximations. A replacement of the second-order derivative with the highly accurate five-point stencil,
\begin{equation}
f''(x) \approx \frac{-f(x+2\Delta x) + 16f(x+\Delta x) - 30f(x) + 16f(x-\Delta x) - f(x-2\Delta x)}{12\Delta x^2} + \mathcal{O}(\Delta x^4) ,
\end{equation}
implies that Eq.~\eqref{eq:ImpExpExpression} is now discretized as
\begin{equation}
\begin{split}
\psi_j^{n+1} + \frac{i \Delta t}{2\hbar} \qty[ -\frac{\hbar^2}{2m} \qty(  \frac{-\psi_{j+2}^{n+1} +16\psi_{j+1}^{n+1} - 30\psi_{j}^{n+1} +16\psi_{j-1}^{n+1} -\psi_{j-2}^{n+1}}{12\Delta x^2}  ) + V_j  \psi_j^{n+1}  ] \\ = \psi_j^{n} - \frac{i \Delta t}{2\hbar} \qty[ -\frac{\hbar^2}{2m} \qty( \frac{-\psi_{j+2}^{n} +16\psi_{j+1}^{n} - 30\psi_{j}^{n} +16\psi_{j-1}^{n} -\psi_{j-2}^{n}}{12\Delta x^2} )  + V_j  \psi_j^{n}  ].
\end{split}
\end{equation}
Following a similar approach as in the previous section, we denote
\begin{eqnarray}
\zeta_j^n &=& \psi_j^{n} - \frac{i \Delta t}{2\hbar} \qty[ -\frac{\hbar^2}{2m} \qty(  \frac{-\psi_{j+2}^{n} +16\psi_{j+1}^{n} - 30\psi_{j}^{n} +16\psi_{j-1}^{n} -\psi_{j-2}^{n}}{12\Delta x^2}  )  + V_j  \psi_j^{n} ],
\nonumber	\\
a_j &=& 1 + \frac{i\Delta t}{2\hbar} \qty( \frac{5\hbar^2}{4m\Delta x^2} + V_j  ),
\hspace{1cm}
b = -\frac{i\hbar\Delta t}{3m\Delta x^2},
\hspace{1cm}
c = \frac{i\hbar\Delta t}{48m\Delta x^2},
\end{eqnarray}
which reduces the problem to
\begin{equation}
\begin{pmatrix}
a_1 & b & c \\
\ddots & \ddots & \ddots & \ddots \\
& \ddots & \ddots & \ddots & \ddots & \ddots \\
& & c & b & a_{j-1} & b & c
\\ & & & c & b & a_j & b & c
\\ & & & & c & b & a_{j+1} & b & c
\\ & & & & & \ddots & \ddots & \ddots & \ddots & \ddots
\\ & & & & & & \ddots & \ddots & \ddots & \ddots
\\ & & & & & & & c & b & a_{J-2}
\end{pmatrix} \\
\cdot \begin{pmatrix}
\psi_1^{n+1} \\
\vdots \\
\psi_{j-2}^{n+1} \\
\psi_{j-1}^{n+1} \\
\psi_{j}^{n+1} \\
\psi_{j+1}^{n+1} \\
\psi_{j+2}^{n+1} \\
\vdots\\
\psi_{J-2}^{n+1} \end{pmatrix} = \begin{pmatrix}
\zeta_1^{n} \\
\vdots \\
\zeta_{j-2}^{n} \\
\zeta_{j-1}^{n} \\
\zeta_{j}^{n} \\
\zeta_{j+1}^{n} \\
\zeta_{j+2}^{n} \\
\vdots \\
\zeta_{J-2}^{n}
\end{pmatrix}.
\label{eq:PDiagEqn}
\end{equation}
We now have a pentadiagonal system of linear equations for $J-2$ unknown wave function values at time $t_{n+1}$. The Thomas algorithm is not applicable anymore, as it works only for tridiagonal matrices. Accordingly, we chose the LU-factorisation techniques that are versatile enough to solve both the tridiagonal and the pentadiagonal system of equations. We perform an LU-factorisation of the constant matrix on the left, followed by forward and backward substitutions of the $\zeta$ vector on the right~\cite{CoP_PPuschnig,Ankit_2021_Quantum}. A Python implementation is publicly available at Zenodo and GitHub~\cite{Ankit_TDSE_Zenodo,Ankit_TDSE_GitHub}, with the corresponding documentation in Ref.~\cite{Ankit_2022_TDSE}. Note that these methods work only for square-integrable wave functions, which are well localised all the time. 
Otherwise, the wave function may get reflected from the numerical boundaries, leading to unwanted interference.

\subsection{Comparison of numerical errors}

In this section we calculate the evolution of a Gaussian wave packet with the standard tridiagonal method and the pentadiagonal method developed in this work. The wave packet is initially centered around $x_0$ with a position spread $\sigma$ and a momentum $p_0$:
\begin{equation}
\psi(x,t=0) = \frac{1}{\sqrt{\sigma\sqrt{2\pi}}}
\ \exp\qty( -\frac{(x-x_0)^2}{4\sigma^2} + i \frac{p_0}{\hbar}(x-x_0) ) .
\end{equation}
We compare the numerical errors accumulated in both methods for the case of evolution in free space and the harmonic oscillator potential.
As a parameter of interest, we chose Heisenberg's uncertainty product $\bm{\Delta} x \bm{\Delta} p$, since it involves both the statistical moments of the position and momentum variables.

\begin{enumerate}[label=\alph*)]

\item Evolution in the free space, i.e., $V = 0$, can be solved analytically using Fourier transformation techniques~\cite{GaussEvolFreeSpace_SMBlinder}:
\begin{equation}
\psi(x,t) = \frac{1}{\sqrt{\sigma(1+i\omega_0 t)\sqrt{2\pi}}} \ \exp[ - \frac{1}{4\sigma^2(1 + i \omega_0 t)} \qty( x-x_0-2i \sigma^2 \frac{p_0}{\hbar} )^2 - \sigma^2\frac{p_0^2}{\hbar^2}  ],
\end{equation}
where $\omega_0 = \hbar/2m\sigma^2$. This implies that the spread in the position and the momentum spaces are given by
\begin{equation}
\bm{\Delta}x = \sqrt{ \ev{\hat x^2} - \ev{\hat x}^2 } = \sigma\sqrt{1+\omega_0^2t^2},
\hspace{1cm}
 \bm{\Delta}p  = \sqrt{ \ev{\hat p^2} - \ev{\hat p}^2 } = \frac{\hbar}{2\sigma},
\end{equation}
Hence, the uncertainty product is
\begin{equation}
\bm{\Delta}x  \bm{\Delta}p = \frac{\hbar}{2}\sqrt{1+\omega_0^2t^2}.
\label{eq:uncp_FS}
\end{equation}

\item Evolution in the harmonic oscillator potential, i.e., $V = \frac{1}{2}m\omega^2x^2$, can also be solved analytically~\cite{GaussEvolHarmOsc_Tsuru}. The closed form is rather complicated, but it gives the position and momentum spreads as
\begin{equation}
	\bm{\Delta} x   =  \sigma \sqrt{ \cos[2](\omega t) + \frac{\omega_0^2}{\omega^2} \sin[2](\omega t) },
\hspace{1cm}
	\bm{\Delta} p  =  \frac{\hbar}{2\sigma} \sqrt{ \cos[2](\omega t) + \frac{\omega^2}{\omega_0^2} \sin[2](\omega t) },
\end{equation}
The corresponding uncertainty product is given by
\begin{equation}
\bm{\Delta} x \bm{\Delta} p  = \frac{\hbar}{2}\sqrt{ \cos^4(\omega t) + \sin^4(\omega t) + \frac{1}{4} \qty(  \frac{\omega_0^2}{\omega^2} + \frac{\omega^2}{\omega_0^2}  ) \sin^2(2\omega t) },
\label{eq:uncp_HO}
\end{equation}

\end{enumerate}

\begin{figure}[!t] 
\centering

\subfloat[Evolution in the free space, $V = 0$. Initial wave packet is centered at $x = -50$ with a width of $2$ units and a momentum of $+1$ unit.]{\includegraphics[width=\linewidth]{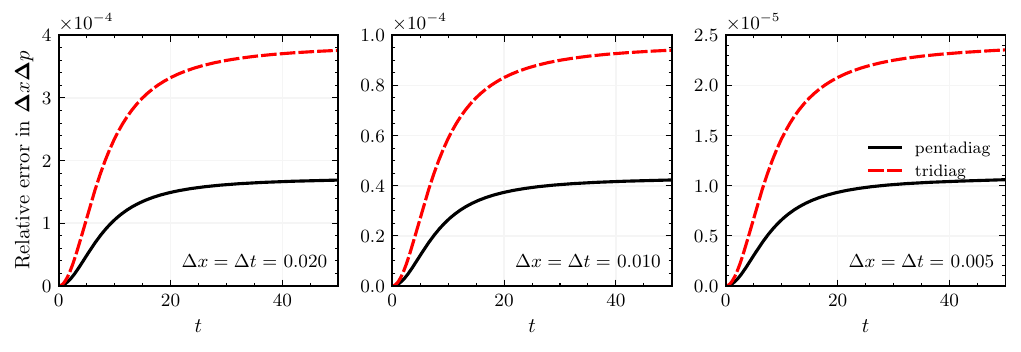}\label{fig:errors_cayley_FS}}

\subfloat[Evolution in the harmonic oscillator potential, $V=\frac{1}{2}m\omega^2x^2$ with $\omega = 0.1$. Initial wave packet is centered at $x = -10$, with a width of $2$ units.]{\includegraphics[width=\linewidth]{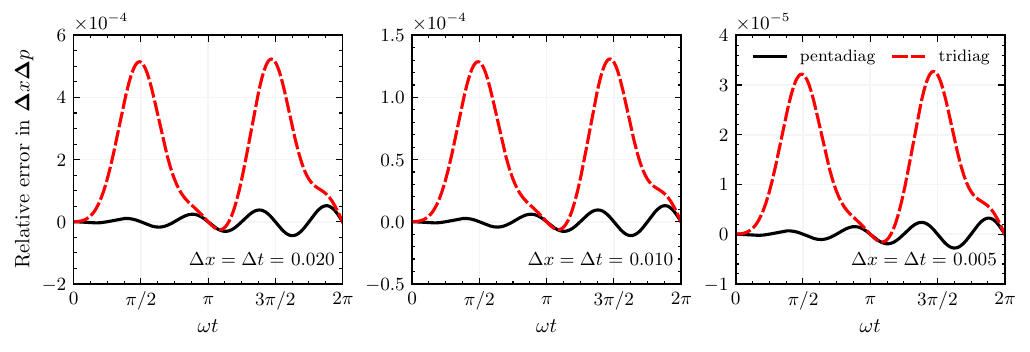}\label{fig:errors_cayley_HO}}

\caption{Comparison of errors in the tridiagonal and the pentadiagonal solutions of the time-dependent Schr\"{o}dinger equation. $\bm{\Delta} x \bm{\Delta} p$ is the Heisenberg's uncertainty product. We assume $\hbar = 1$, $m = 1$, and the relative errors are calculated w.r.t. the analytical results discussed in the main text. $\Delta x$ denotes the grid size, and $\Delta t$ is the time step. Note different vertical scales in each panel.}

\label{fig:errors_cayley}

\end{figure}

Assuming $\hbar = 1$, $m = 1$, we calculate the time evolution with both the tridiagonal and the pentadiagonal methods. In Fig.~\ref{fig:errors_cayley} we show the errors in the uncertainty product, calculated w.r.t. the closed forms in Eqs.~\eqref{eq:uncp_FS} and~\eqref{eq:uncp_HO}. It can be easily seen that our pentadiagonal solutions are far more accurate than the standard ones. Accordingly, they will be used for simulating quantum evolution in extremely weak fields. 
We have used the standard tridiagonal solutions for studying the head-on collision of charged particles~\cite{Ankit_2021_Quantum}, and the highly accurate pentadiagonal solutions for the astonishingly weak gravitational coupling between two nearby quantum objects~\cite{Ankit_2022_Gravity}.

Note that Heisenberg's uncertainty product requires the first two statistical moments of position and momentum operators, which can be evaluated with:
\begin{equation}
    \ev{\hat x^n}  =  \int_{-\infty}^{+\infty} dx \ \psi^* \ x^n \ \psi,
\hspace{1cm}
    \ev{\hat p^n}  =  (-i\hbar)^n \int_{-\infty}^{+\infty} dx \ \psi^* \ \pdv[n]{\psi}{x}.
\end{equation}
It should be noted that $\ev{\hat p^2}$ can be calculated without the additional evaluation of the second-order derivative.
For a well localised problem, the wave function is square integrable: $\lim_{x\to\pm \infty} \psi = 0$ and $\lim_{x\to\pm \infty} d\psi/dx = 0$, and integration by parts implies
\begin{equation}
\ev{\hat p^2} 
=
 -\hbar^2 \int_{-\infty}^{+\infty} dx \ \psi^* \pdv[2]{\psi}{x}
=
-\hbar^2 \qty[ \psi^* \pdv{\psi}{x} - \int dx \ \pdv{\psi^*}{x} \pdv{\psi}{x} ]_{-\infty}^{+\infty}
=
\hbar^2 \int_{-\infty}^{+\infty} dx \ \abs{\pdv{\psi}{x}}^2.
\end{equation}
Furthermore, by utilizing the law of conservation of energy, we calculate $\ev{\hat p^2}$ without involving any numerical derivative whatsoever.
A unitary evolution implies that the total energy, $\ev{\hat H}$, is a constant of motion. At $t=0$ we start with a minimum uncertainty Gaussian wave packet characterized by $\bm{\Delta} x \bm{\Delta} p (0) = \hbar/2$, which implies
\begin{equation}
\ev{\hat p^2(0)} = \ev{\hat p(0)}^2 + \bm{\Delta} p^2 (0) = p_0^2 + \frac{\hbar^2}{4\sigma^2}.
\end{equation}
On equating $\ev{\hat H(0)}$ with $\ev{\hat H}$ we arrive at
\begin{equation}
    \ev{\hat p^2} = p_0^2 + \frac{\hbar^2}{4\sigma^2} + 2m \Big(  \ev{V(0)} - \ev{V} \Big) ,
\end{equation}
where $\ev{V(0)}$ is readily available in closed formulas, e.g., in the free space $\ev{V(0)} = 0$, and in the harmonic oscillator potential $\ev{V(0)} = \frac{1}{2}m\omega_0^2\ev{\hat x^2(0)} = \frac{1}{2}m\omega_0^2(x_0^2+\sigma^2)$.

\subsection{The heartbeating inside a box}

\begin{figure}[!t]
\centering
\includegraphics[width=\linewidth]{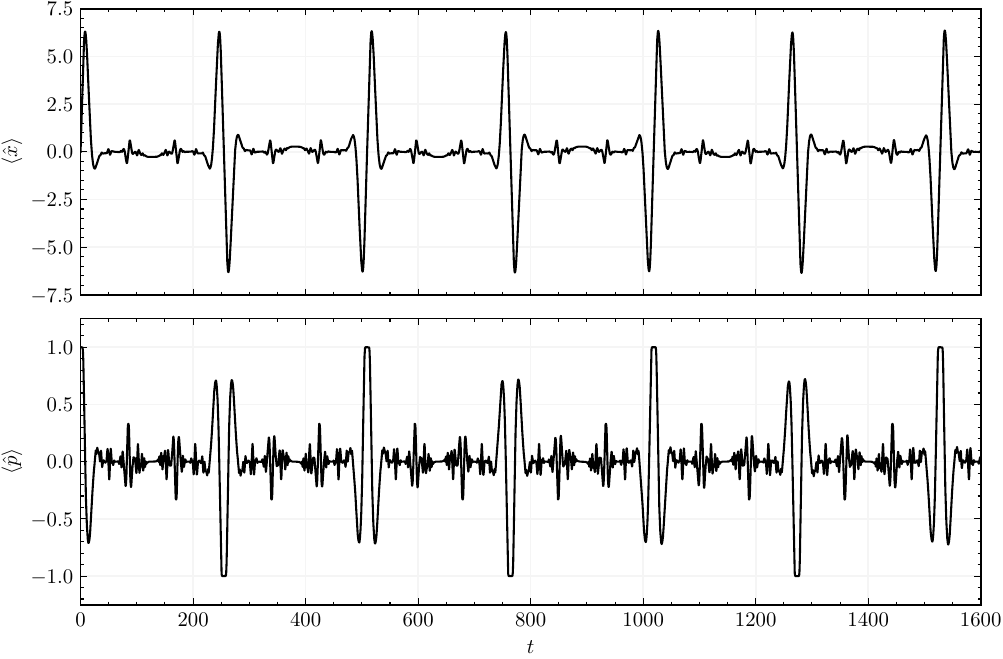}
\caption{Expected position and momentum of a Gaussian evolving inside a box extending from $x=-10$ to $+10$. Assuming $\hbar = m = 1$, the initial wave packet is centered at the origin with a width of 1 unit and a momentum of +1 unit. The size of the position grid is $\Delta x = 0.01$, and the time step is $\Delta t = 0.01$.}
\label{fig:heartbeat_in_a_box}
\end{figure}

Now that we have a numerical scheme for resolving the quantum dynamics of localised wave packets, in this section we play around and demonstrate the fascinating dance of an (initially Gaussian) wave packet evolving inside a box. Assuming $\hbar  = 1$, we consider a particle of mass $m = 1$ trapped inside a box extending between $x = \pm 10$. As a sanity check, it is first confirmed that the wave function does not change with time when the initial state corresponds to any one of the eigenstates:
\begin{equation}
\psi_n(x,0) = \begin{cases}
\sqrt{\frac{2}{L}} \sin \qty( \frac{n\pi x}{L} ) ,	\hspace{1cm}	n = 0,2,4,\dots 
\\
\sqrt{\frac{2}{L}} \cos \qty( \frac{n\pi x}{L} ) ,	\hspace{1cm}	n = 1,3,5,\dots 
\end{cases}
\end{equation}
We then consider the initial state as a Gaussian wave packet with a width of $1$ unit centered at the origin with $1$ unit of momentum to the right. This way, the forward part of the wave packet hits the boundary at $x = +10$ and is reflected back. These reflections interfere with the rest of the wave packet to create a beautiful dancing pattern.
In Fig.~\ref{fig:heartbeat_in_a_box} we show the expected position and momentum as a function of time which looks like a periodically repeating heartbeat pattern.

\section{Bipartite wave packet dynamics}	
\label{appendix:TDSE-Transf2COM}

Till now we have discussed the case of a single-particle wave packet evolving in a classical background potential.
It turns out that the same methods can be utilised to solve the bipartite dynamics after a careful change of coordinates. The time-dependent Schr\"{o}dinger equation (TDSE) for a system of two particles $A$ and $B$ is given by
\begin{eqnarray}
	\Bigg( -\frac{\hbar^2}{2m_A} \pdv[2]{x_A}  -\frac{\hbar^2}{2m_B} \pdv[2]{x_B} + V(\hat x_A,\hat x_B) \Bigg) \Psi(x_A,x_B,t) 
	= i\hbar\pdv{t} \Psi(x_A,x_B,t),
\end{eqnarray}
where $x_A$ and $x_B$ are the positions/displacements of the masses $m_A$ and $m_B$, and $\hat p_A = -i\hbar \partial/\partial x_A$ and $\hat p_B = -i\hbar \partial/\partial x_B$ are their respective momenta. In an attempt to decouple this two-body problem, we make a coordinate transformation to the COM frame of reference:
\begin{equation}
R = \frac{m_Ax_A+m_Bx_B}{m_A+m_B},
\hspace{1cm}
r = x_B-x_A,
\end{equation}
where $R$ and $r$ are the positions/displacements of the COM [mass $M = m_A+m_B$] and the reduced mass [mass $\mu = m_Am_B/(m_A+m_B)$], respectively. One can take the time derivatives to write their respective momenta as
\begin{eqnarray}
P = M\dv{R}{t} 	= \frac{m_A+m_B}{m_A+m_B} \qty( m_A \dv{x_A}{t} + m_B \dv{x_B}{t} ) 	= p_A+p_B,
\nonumber	\\
p = \mu\dv{r}{t} = \frac{m_Am_B}{m_A+m_B} 	\qty( \dv{x_B}{t} - \dv{x_A}{t} )	 = \frac{m_Ap_B-m_Bp_A}{m_A+m_B} ,
\end{eqnarray}
which implies that the inverse transformations are
\begin{equation}
	x_A = R  -  \frac{m_B}{M}r,
\hspace{1cm}
	x_B = R + \frac{m_A}{M}r,
\hspace{1cm}
	p_A = \frac{m_A}{M}P - p,
\hspace{1cm}
	p_B = \frac{m_B}{M}P + p.
\end{equation}
The displacements $x_A$ and $x_B$ are functions of $R$ and $r$, and the rules of differentiation imply
\begin{eqnarray}
\pdv{x_A} &=& \qty( \pdv{R}{x_A} ) \pdv{R}		+	\qty( \pdv{r}{x_A} ) \pdv{r}	= \frac{m_A}{M} \pdv{R} - \pdv{r},
\\
\pdv{x_B} &=& \qty( \pdv{R}{x_B} ) \pdv{R}		+	\qty( \pdv{r}{x_B} ) \pdv{r}	= \frac{m_B}{M} \pdv{R} + \pdv{r}.
\end{eqnarray}
Similarly, the second-order derivatives can be calculated as
\begin{eqnarray}
\pdv[2]{x_A} &=& \qty( \pdv{R}{x_A} ) \pdv{R}	 \qty( \frac{m_A}{M} \pdv{R} - \pdv{r} ) + \qty( \pdv{r}{x_A} ) \pdv{r}	 \qty( \frac{m_A}{M} \pdv{R} - \pdv{r} )	\nonumber \\
&& =	\frac{m_A}{M} \qty( \frac{m_A}{M} \pdv[2]{R} - \pdv{}{R}{r} )	- 	\qty( \frac{m_A}{M} \pdv{}{r}{R} - \pdv[2]{r} )	\nonumber	\\
&& =  \frac{m_A^2}{M^2} \pdv[2]{R} + \pdv[2]{r} - 2\frac{m_A}{M}\pdv{}{R}{r},
\\	\nonumber \\
\pdv[2]{x_B} &=& \qty( \pdv{R}{x_B} ) \pdv{R}	 \qty( \frac{m_B}{M} \pdv{R} + \pdv{r} ) + \qty( \pdv{r}{x_B} ) \pdv{r}	 \qty( \frac{m_B}{M} \pdv{R} + \pdv{r} )	\nonumber \\
&& =	\frac{m_B}{M} \qty( \frac{m_B}{M} \pdv[2]{R} + \pdv{}{R}{r} )	+ 	\qty( \frac{m_B}{M} \pdv{}{r}{R} + \pdv[2]{r} )	\nonumber	\\
&& =  \frac{m_B^2}{M^2} \pdv[2]{R} + \pdv[2]{r} + 2\frac{m_B}{M}\pdv{}{R}{r}.
\end{eqnarray}
which implies that the kinetic energy part of the Hamiltonian transforms as
\begin{eqnarray}
- \frac{\hbar^2}{2m_A} \pdv[2]{x_A}  &-& \frac{\hbar^2}{2m_B} \pdv[2]{x_B} 
\nonumber	\\
&=& -\frac{\hbar^2}{2m_A} \qty( \frac{m_A^2}{M^2} \pdv[2]{R} + \pdv[2]{r} - 2\frac{m_A}{M}\pdv{}{R}{r} )	 - 	\frac{\hbar^2}{2m_B} \qty( \frac{m_B^2}{M^2} \pdv[2]{R} + \pdv[2]{r} + 2\frac{m_B}{M}\pdv{}{R}{r} )	
\nonumber	\\
&=& -\frac{\hbar^2}{2} \qty(\frac{m_A+m_B}{M^2}) \pdv[2]{R} - \frac{\hbar^2}{2} \qty(\frac{1}{m_A}+\frac{1}{m_B}) \pdv[2]{r}	
\nonumber \\
&=& -\frac{\hbar^2}{2M} \pdv[2]{R}  -\frac{\hbar^2}{2\mu} \pdv[2]{r}.
\end{eqnarray}
Within the scope of this thesis we deal only with central interactions, i.e., the potential is a function of the relative separation only: $V(x_A,x_B) = V(x_B-x_A)=V(r)$, and hence the Schr\"odinger equation looks like
\begin{equation}
	\Bigg( -\frac{\hbar^2}{2M} \pdv[2]{R}   -\frac{\hbar^2}{2\mu} \pdv[2]{r} + V(\hat r) \Bigg) \Psi(x_A,x_B,t)
	=  i\hbar\pdv{t}\Psi(x_A,x_B,t).
	\label{eq:TDSE_in_COM_general}
\end{equation}
Given that the initial wave function transforms to the COM frame as $\Psi(x_A,x_B,t=0) = \phi(R,t=0) \ \psi(r,t=0)$, the separation of variables in Eq.~\eqref{eq:TDSE_in_COM_general} will ensure that the product form is maintained at all times. The problem will now decouple as
\begin{equation}
	-\frac{\hbar^2}{2M} \pdv[2]{R} \phi(R,t)  =  i\hbar\pdv{t}\phi(R,t),	
\label{eq:COMSE-free}
\end{equation}
\begin{equation}
	\qty( -\frac{\hbar^2}{2\mu} \pdv[2]{r} + V(\hat r) ) \psi(r,t) = i\hbar\pdv{t}\psi(r,t),
\end{equation}
where $\hat P = -i\hbar\partial/\partial R$ and $\hat p = -i\hbar\partial/\partial r$ can now be identified as the momentum operators for the COM and the reduced mass, respectively. Here the COM evolves in the free space, which can be easily solved with analytical techniques~\cite{GaussEvolFreeSpace_SMBlinder}. The reduced mass evolves under the influence of the interaction $V(r)$, and depending on its functional form, one can use either the analytical or the numerical method to calculate the corresponding time evolution. The two-body wave function is given by the product
\begin{equation}
	\Psi(x_A,x_B,t) = \phi \qty( \frac{m_Ax_A+m_Bx_B}{m_A+m_B},t  ) \ \psi \qty( x_B-x_A,t ).
	\label{eq:TBWF_LAB2COM_relation}
\end{equation}

\subsection{Transformation of a two-mode Gaussian state}
\label{sec:transform_TMGS}

\begin{figure}
	\centering
	\includegraphics[width=\linewidth]{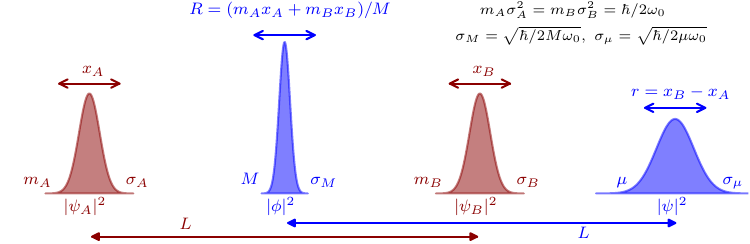}
	\caption{From LAB frame to COM frame. Gaussianity of the initial state is preserved as well as the product form. The widths, however, are different in different frames.}
	\label{fig:TMGS_genreal}
\end{figure}

Recall that in the previous section the bipartite TDSE decouples to two independent TDSEs only when the initial state can be written as a product form in the COM frame of reference. In this section we derive the conditions under which this happens for a two-mode Gaussian state. For later convenience in the two-body problems, $x_A$ and $x_B$ shall denote the displacements of the two masses from their initial positions.
The initial state describing two independent masses prepared in Gaussian states with position spreads $\sigma_A$ and $\sigma_B$ is $\Psi(x_A,x_B,t=0) = \psi_A(x_A) \ \psi_B(x_B)$, with
\begin{eqnarray}
\psi_A(x_A) &=& \qty( \frac{1}{2\pi\sigma_A^2} )^{1/4} \exp(-\frac{x_A^2}{4\sigma_A^2} + i\frac{p_{A0} }{\hbar}x_A ),
\\
\psi_B(x_B) &=& \qty( \frac{1}{2\pi\sigma_B^2} )^{1/4} \exp(-\frac{x_B^2}{4\sigma_B^2} + i\frac{p_{B0} }{\hbar}x_B ) ,
\end{eqnarray}
where $p_{A0}$ and $p_{B0}$ denote the initial momenta of the two particles. With simple algebra, we can rearrange the initial wave function as
\begin{eqnarray}
\Psi(t=0) &=& \qty( \frac{1}{2\pi\sigma_A^2} )^{1/4} \exp(-\frac{x_A^2}{4\sigma_A^2} + i\frac{p_{A0} }{\hbar}x_A ) \ \qty( \frac{1}{2\pi\sigma_B^2} )^{1/4} \exp(-\frac{x_B^2}{4\sigma_B^2} + i\frac{p_{B0} }{\hbar}x_B )	\nonumber	\\
&=&   \qty( \frac{1}{2\pi\sigma_A^2})^{1/4} \qty( \frac{1}{2\pi\sigma_B^2} )^{1/4} \exp(-\frac{x_A^2}{4\sigma_A^2}  -\frac{x_B^2}{4\sigma_B^2}) \exp\qty( i\frac{ p_{A0}x_A+p_{B0}x_B }{\hbar} ) .	
\end{eqnarray}
We shall now make use of the inverse transformations to express this in the COM frame. To start with, the Gaussian part is given by
\begin{eqnarray}
\frac{x_A^2}{\sigma_A^2} + \frac{x_B^2}{\sigma_B^2} &=& \frac{1}{\sigma_A^2} \qty( R - \frac{m_B}{M}r )^2 + \frac{1}{\sigma_B^2} \qty( R + \frac{m_A}{M}r )^2
\nonumber	\\
&=& \frac{ 1 }{\sigma_A^2} \qty( R^2 + \frac{m_B^2}{M^2}r^2 - 2\frac{m_B}{M}Rr  ) + \frac{ 1 }{\sigma_B^2} \qty( R^2 + \frac{m_A^2}{M^2}r^2 + 2\frac{m_A}{M}Rr )
\nonumber	\\
&=&	\qty( \frac{ \sigma_A^2+\sigma_B^2 }{\sigma_A^2\sigma_B^2} ) R^2 + \qty( \frac{m_A^2\sigma_A^2+m_B^2\sigma_B^2}{M^2\sigma_A^2\sigma_B^2}) r^2 + \qty( \frac{m_A\sigma_A^2-m_B\sigma_B^2}{M\sigma_A^2\sigma_B^2}) Rr.
\end{eqnarray}
The last term needs to vanish for the state to decouple into independent Gaussians in $R$ and $r$, which happens only when $m_A\sigma_A^2=m_B\sigma_B^2$. Note that
\begin{equation}
m_A \sigma_A^2 = m_A \times \frac{\hbar}{2m_A\omega_A} = \frac{\hbar}{2\omega_A},
\hspace{1cm}
m_B \sigma_B^2 = m_B \times \frac{\hbar}{2m_B\omega_B} = \frac{\hbar}{2\omega_B},
\end{equation}
and hence the conditionality $m_A\sigma_A^2=m_B\sigma_B^2$ essentially implies $\omega_A = \omega_B \equiv \omega_0$. The wave packet dynamics decouples only when the two particles are prepared in the ground state of identical harmonic traps of frequency $\omega_0$. Under this assumption,
\begin{eqnarray}
\frac{x_A^2}{\sigma_A^2} + \frac{x_B^2}{\sigma_B^2} &=&	\qty[ \frac{ \qty(\frac{\hbar}{2m_A\omega_0}) + \qty(\frac{\hbar}{2m_B\omega_0}) } {  \qty(\frac{\hbar}{2m_A\omega_0}) \qty(\frac{\hbar}{2m_B\omega_0}) } ] R^2 + \qty[ \frac{m_A^2\qty(\frac{\hbar}{2m_A\omega_0})+m_B^2\qty(\frac{\hbar}{2m_B\omega_0})}{M^2\qty(\frac{\hbar}{2m_A\omega_0})\qty(\frac{\hbar}{2m_B\omega_0})} ] r^2 
\nonumber	\\
&=& \qty[ \frac{2(m_A+m_B)\omega_0}{\hbar} ] R^2 +  \qty[ \frac{2m_Am_B\omega_0}{(m_A+m_B)\hbar} ] r^2
\nonumber	\\
&=& \qty( \frac{2M\omega_0}{\hbar} ) R^2 +  \qty( \frac{2\mu\omega_0}{\hbar} ) r^2
\nonumber	\\
&=&	\frac{R^2}{\sigma_M^2} + \frac{r^2}{\sigma_\mu^2},
\end{eqnarray}
where $\sigma_M^2 = \hbar/2M\omega_0$ and $\sigma_\mu^2 = \hbar/2\mu\omega_0$. For the plane wave part of $\Psi(t=0)$ we have
\begin{eqnarray}
p_{A0}x_A+p_{B0}x_B &=& p_{A0}\qty( R - \frac{m_B}{M}r ) + p_{B0}\qty( R + \frac{m_A}{M}r )	\nonumber	\\
&=&	\qty(p_{A0}+p_{B0})R + \qty(\frac{m_Ap_{B0}-m_Bp_{A0}}{m_A+m_B}) r	\nonumber	\\
&=& p_{M0}R + p_{\mu 0}r ,
\end{eqnarray}
where $p_{M0}$ and $p_{\mu 0}$ correspond to the total initial momenta for the COM and the reduced mass, respectively.
At last, in the normalisation constant we can put
\begin{eqnarray}
\sigma_A^2 \sigma_B^2 	&=&	\frac{\hbar}{2m_A\omega_0} \times \frac{\hbar}{2m_B\omega_0}
\nonumber	\\
&=&		\frac{\hbar}{2\omega_0} \times \frac{m_A+m_B}{m_A m_B} \times \frac{1}{m_A+m_B} \times \frac{\hbar}{2\omega_0} 
\nonumber	\\
&=&		\frac{\hbar}{2 M \omega_0} \times \frac{\hbar}{2 \mu \omega_0}
\nonumber	\\
&\equiv&		\sigma_M^2 \sigma_\mu^2 .
\end{eqnarray}
With all these transformations, the initial wave function nicely separates as
\begin{eqnarray}
\Psi(t=0) &=& \qty( \frac{1}{2\pi\sigma_M^2} )^{1/4} \qty( \frac{1}{2\pi\sigma_\mu^2} )^{1/4} \exp( -\frac{R^2}{4\sigma_M^2} - \frac{r^2}{4\sigma_\mu^2} ) \exp(i\frac{p_{M0}R + p_{\mu 0}r}{\hbar})	\nonumber	\\
&=& \qty( \frac{1}{2\pi\sigma_M^2} )^{1/4} \exp(-\frac{R^2}{4\sigma_M^2} + i \frac{p_{M0}}{\hbar}R) \ \qty( \frac{1}{2\pi\sigma_\mu^2} )^{1/4} \exp(-\frac{r^2}{4\sigma_\mu^2} + i\frac{p_{\mu 0}}{\hbar}r )	\nonumber	\\
&=& \phi(R,t=0) \ \psi(r,t=0) ,
\label{eq:InitialState_inCOMframe_generalised}
\end{eqnarray}
where $\phi(R,t=0)$ and $\psi(r,t=0)$ describe the initial states for the COM and the reduced mass, respectively:
\begin{eqnarray}
\phi(R,t=0) &=& \qty( \frac{1}{2\pi\sigma_M^2} )^{1/4} \exp(-\frac{R^2}{4\sigma_M^2} + i \frac{p_{M0}}{\hbar}R) ,
\\
\psi(r,t=0) &=& \qty( \frac{1}{2\pi\sigma_\mu^2} )^{1/4} \exp(-\frac{r^2}{4\sigma_\mu^2} + i\frac{p_{\mu 0}}{\hbar}r ) .
\label{eq:COMwavefunction_t0_general}
\end{eqnarray}

The COM wave packet admits a width of $\sigma_M = \sqrt{\hbar/2M\omega_0}$, and the reduced mass wave packet has a width of $\sigma_\mu = \sqrt{\hbar/2\mu\omega_0}$. For two identical masses $m$ prepared in Gaussians of width $\sigma$, the COM would have a smaller width of $\sigma/\sqrt{2}$, and the reduced mass will have a larger width of $\sigma\sqrt{2}$. The corresponding relations are illustrated in Fig.~\ref{fig:TMGS_genreal}. A separable Hamiltonian implies that the two-body wave function retains its product form at all times, i.e., $\Psi(x_A,x_B,t) = \phi(R,t) \ \psi(r,t)$. 
Note that the time-dependence of $\phi$ is governed by Eq.~\eqref{eq:COMSE-free}, and is solvable analytically~\cite{GaussEvolFreeSpace_SMBlinder}:
\begin{equation}
\phi(R,t) = \frac{1}{\sqrt{\sigma_M(1+i\omega_0 t)\sqrt{2\pi}}} \ \exp[ - \frac{1}{4\sigma_M^2(1 + i \omega_0 t)} \qty( R - 2i \sigma_M^2 \frac{p_{M0}}{\hbar} )^2 - \sigma_M^2\frac{p_{M0}^2}{\hbar^2}  ] .
\label{eq:TDWF_COM_general_SMBlinder}
\end{equation}

\subsection{The case of optomechanically held masses}

In most physical problems the COM is a free particle. However, there might be situations where this is not true, e.g., the when the two particles are always trapped in harmonic potentials:
\begin{equation}
\hat H = -\frac{\hbar^2}{2m_A} \pdv[2]{x_A} + \frac{1}{2}m_A\omega_0^2\hat x_A^2 -\frac{\hbar^2}{2m_B} \pdv[2]{x_B} + \frac{1}{2}m_B\omega_0^2\hat x_B^2 + V(\hat x_B-\hat x_A).
\end{equation}
We can use inverse coordinate transformations to prove that
\begin{eqnarray}
\frac{1}{2}m_A\omega_0^2x_A^2 &+& \frac{1}{2}m_B\omega_0^2x_B^2 
\nonumber	\\
&=&  \frac{1}{2}m_A\omega_0^2 \qty(R - \frac{m_B}{M}r)^2  +   \frac{1}{2}m_B\omega_0^2 \qty(R + \frac{m_A}{M}r)^2
\nonumber	\\
&=&		\frac{1}{2}m_A\omega_0^2 \qty( R^2 + \frac{m_B^2}{M^2}r^2 - 2\frac{m_B}{M}Rr )
	+ \frac{1}{2}m_B\omega_0^2 \qty( R^2 + \frac{m_A^2}{M^2}r^2 + 2\frac{m_A}{M}Rr )
\nonumber	\\
&=& 	\frac{1}{2} (m_A+m_B) \omega_0^2 R^2 + \frac{1}{2} \qty(\frac{m_Am_B}{M}) \omega_0^2 r^2	\nonumber	\\
&=& 	\frac{1}{2} M \omega_0^2 R^2 + \frac{1}{2} \mu \omega_0^2 r^2.
\end{eqnarray}
Similar to the previous section, the bipartite TDSE now decouples into
\begin{eqnarray}
	\qty(	-\frac{\hbar^2}{2M} \pdv[2]{R} + \frac{1}{2} M \omega_0^2 \hat R^2 ) \phi(R,t)  &=&  i\hbar\pdv{t}\phi(R,t),	
\label{eq:COMSE-trap}
\\
	\qty( -\frac{\hbar^2}{2\mu} \pdv[2]{r} + \frac{1}{2} \mu \omega_0^2 \hat r^2 +  V(\hat r) ) \psi(r,t) &=& i\hbar\pdv{t}\psi(r,t),
\end{eqnarray}
which clearly shows the the COM is trapped in a virtual harmonic potential with a trap frequency the same as that for the two particles.

\section{Summary}

We demonstrated the utility of Cayley's form of evolution operator in the numerical resolution of continuous-variable quantum dynamics. The highly accurate five-point stencil was utilized to discretize the problem as an implicit-explicit pentadiagonal Crank-Nicolson scheme, which is unconditionally stable on realistic time scales. Given the same grid size and time step, the resultant numerical solutions achieve much higher accuracy than the standard ones.
We also discussed the coordinate transformations to the COM frame of reference and the situations when the bipartite TDSE decouples into two single-particle TDSEs. For a two-mode Gaussian state, this happens only when the two particles are prepared in the ground state of identical harmonic traps. The theory works for arbitrary central interaction and for multiple central forces acting at the same time.


\begin{thebibliography}{10}

\bibitem{GaussEvolFreeSpace_SMBlinder}
S.~M. Blinder.
\newblock ``{Evolution of a Gaussian Wavepacket}''.
\newblock \href{https://dx.doi.org/10.1119/1.1974961}{American Journal of
  Physics {\bf 36}, 525}~(1968).

\bibitem{GaussEvolHarmOsc_Tsuru}
H.~Tsuru.
\newblock ``{Wave Packet Motion in Harmonic Potential}''.
\newblock \href{https://dx.doi.org/10.1143/JPSJ.60.3657}{Journal of the
  Physical Society of Japan {\bf 60}, 3657}~(1991).

\bibitem{FEIT1982412}
M.~D. Feit, J.~A. Fleck, and A.~Steiger.
\newblock ``{Solution of the Schr\"odinger equation by a spectral method}''.
\newblock \href{https://dx.doi.org/10.1016/0021-9991(82)90091-2}{Journal of
  Computational Physics {\bf 47}, 412}~(1982).

\bibitem{Park1986}
T.~J. Park and J.~C. Light.
\newblock ``{Unitary quantum time evolution by iterative Lanczos reduction}''.
\newblock \href{https://dx.doi.org/10.1063/1.451548}{The Journal of Chemical
  Physics {\bf 85}, 5870--5876}~(1986).

\bibitem{BANDRAUK1991428}
A.~D. Bandrauk and H.~Shen.
\newblock ``{Improved exponential split operator method for solving the
  time-dependent Schrödinger equation}''.
\newblock \href{https://dx.doi.org/10.1016/0009-2614(91)90232-X}{Chemical
  Physics Letters {\bf 176}, 428}~(1991).

\bibitem{Muller1999}
H.~G. Muller.
\newblock ``{An efficient propagation scheme for the time-dependent
  Schr\"odinger equation in the velocity gauge}''.
\newblock Laser Physics {\bf 9}, 138~(1999).

\bibitem{Nurhuda1999}
M.~Nurhuda and F.~H.~M. Faisal.
\newblock ``{Numerical solution of time-dependent Schr\"odinger equation for
  multiphoton processes: A matrix iterative method}''.
\newblock \href{https://dx.doi.org/10.1103/PhysRevA.60.3125}{Physical Review A
  {\bf 60}, 3125}~(1999).

\bibitem{Watanabe2000}
N.~Watanabe and M.~Tsukada.
\newblock ``{Fast and Stable Method for Simulating Quantum Electron
  Dynamics}''.
\newblock \href{https://dx.doi.org/10.1143/PTPS.138.115}{Progress of
  Theoretical Physics Supplement {\bf 138}, 115}~(2000).

\bibitem{book_FiniteDiff_JWThomas}
J.~W. Thomas.
\newblock ``{Numerical Partial Differential Equations: Finite Difference
  Methods}''.
\newblock Springer-Verlag, USA. ~(1995).
\newblock 1st edition.

\bibitem{KOSLOFF198335}
D.~Kosloff and R.~Kosloff.
\newblock ``{A fourier method solution for the time dependent Schrödinger
  equation as a tool in molecular dynamics}''.
\newblock \href{https://dx.doi.org/10.1016/0021-9991(83)90015-3}{Journal of
  Computational Physics {\bf 52}, 35}~(1983).

\bibitem{Ankit_2022_TDSE}
A.~{Kumar} and P.~{Arumugam}.
\newblock ``{An accurate pentadiagonal matrix solution for the time-dependent
  Schr{\"o}dinger equation}''~(2022).

\bibitem{Ankit_2022_Gravity}
A.~Kumar, T.~Krisnanda, P.~Arumugam, and T.~Paterek.
\newblock ``{Continuous-{V}ariable {E}ntanglement through {C}entral {F}orces:
  {A}pplication to {G}ravity between {Q}uantum {M}asses}''.
\newblock \href{https://dx.doi.org/10.22331/q-2023-05-15-1008}{{Quantum} {\bf
  7}, 1008}~(2023).

\bibitem{Ankit_2021_Quantum}
A.~Kumar, T.~Krisnanda, P.~Arumugam, and T.~Paterek.
\newblock ``{Nonclassical trajectories in head-on collisions}''.
\newblock \href{https://dx.doi.org/10.22331/q-2021-07-19-506}{{Quantum} {\bf
  5}, 506}~(2021).

\bibitem{CoP_PPuschnig}
P.~Puschnig.
\newblock ``{Computerorientierte Physik}''~(2016).

\bibitem{Ankit_TDSE_Zenodo}
A.~Kumar.
\newblock ``{vyason/Cayley-TDSE: An accurate pentadiagonal matrix solution for
  the time-dependent Schr\"{o}dinger equation}''.
\newblock
  \href{https://dx.doi.org/10.5281/zenodo.7275668}{Zenodo:7275668}~(2022).

\bibitem{Ankit_TDSE_GitHub}
{Ankit Kumar}.
\newblock
  ``\href{https://github.com/vyason/Cayley-TDSE}{GitHub/vyason/Cayley-TDSE}''.

\end{thebibliography}

\end{document}